# Stochastic Maximum-Likelihood DOA Estimation and Source Enumeration in the Presence of Nonuniform Noise

Mahmood Karimi

***Abstract*** In this paper, the problem of determining the number of signal sources impinging on an array of sensors and estimating their directions-of-arrival (DOAs) in the presence of spatially white nonuniform noise is considered. It is known that, in the case of nonuniform noise, the stochastic likelihood function cannot be concentrated with respect to the diagonal elements of noise covariance matrix. Therefore, the stochastic maximum-likelihood (SML) DOA estimation and source enumeration in the presence of nonuniform noise requires multidimensional search with very high computational complexity. Recently, two algorithms for estimating noise covariance matrix in the presence of nonuniform noise have been proposed in the literature. Using these new estimates of noise covariance matrix, an approach for obtaining the SML estimate of signal DOAs is proposed. In addition, new approaches are proposed for SML source enumeration with information criteria in the presence of nonuniform noise. The important feature of the proposed SML approaches for DOA estimation and source enumeration is that they have admissible computational complexity. In addition, some of them are robust against correlation between source signals. The performance of the proposed DOA estimation and source enumeration approaches are investigated using computer simulations.

## I. INTRODUCTION

The determination of the number of signals impinging on an array of sensors and estimating their DOAs are fundamental tasks in array signal processing. Several information criteria for source enumeration have been proposed in the last four decades, and in fact, choosing a good penalty term in source enumeration information criteria is still an open problem [1]. Among the most well-known information criteria for source enumeration, the Akaike information criterion (AIC) and the minimum description length (MDL) criterion proposed by Wax and Kailath [2] can be mentioned. Another criterion is the exponentially embedded family (EEF) criterion proposed by Xu and Kay [3]. All these three criteria are based on a complex Gaussian distribution assumption for both source signals and output noises of the array of sensors. Note that the source enumeration criteria like AIC, MDL and EEF need an estimate of the log-likelihood function for each candidate number of sources.

Another group of source enumeration information criteria are based on the model of covariance matrix of the array outputs and the stochastic maximum-likelihood estimator for

M. Karimi is with the School of Electrical and Computer Engineering, Shiraz University, Shiraz, Iran (e-mail: karimi@shirazu.ac.ir).

parameters of this model. The SML-based AIC and MDL criteria proposed by Wax [4] belong to this group. In this approach, the stochastic log-likelihood function is concentrated with respect to both signal covariance matrix and the noise power. This concentration paves the way for estimating the parameters of the likelihood function and the log-likelihood function itself with more admissible computational complexity. The estimated log-likelihood function is then used in the source enumeration criteria like AIC and MDL.

The information criteria that has been so far proposed in the literature for source enumeration usually assume that the sensors' noises are uniform, that is, the power of noise has equal values in all sensors of the array. All criteria mentioned above are derived using this assumption. However, in practice, there are important situations in which the sensors' noises are nonuniform, that is, the power of noise has different values in the sensors of the array [5]-[10]. In these situations, the source enumeration criteria that are based on the uniform noise assumption have poor performance.

Several maximum likelihood methods for DOA estimation in the presence of nonuniform noise have been suggested in the literature. In [10] and [11], the deterministic and stochastic nonuniform maximum likelihood methods, which iteratively estimate noise covariance matrix and DOAs of sources until convergence, are proposed. In [12], an iterative method for estimating signal subspace and noise covariance matrix is proposed. Then, the estimated covariance matrix of noise is used to estimate the DOAs of sources by the deterministic nonuniform maximum likelihood method.

In this paper, we consider the case that the noise of sensors is spatially white nonuniform noise. In this case, the approach of [13] can be used for concentrating the stochastic log-likelihood function with respect to signal covariance matrix. However, more concentration with respect to diagonal elements of noise covariance matrix is not possible in the nonuniform noise case. Therefore, a multidimensional search is required for estimating both directions of arrival (DOA) of sources and diagonal elements of noise covariance matrix. To circumvent this difficulty, we use two estimators of the nonuniform noise covariance matrix which have recently been proposed in [12] and [14]. These estimates for noise covariance matrix make it possible to estimate the DOAs of signal sources and the log-likelihood function with much more admissible computational complexity. In this way, we obtain a new stochastic nonuniform maximum likelihood DOA estimation method that has certain advantages over the existing nonuniform ML DOA estimation methods proposed in the literature. In addition, we use the estimate obtained for the log-likelihood function to propose the stochastic maximum likelihood AIC, MDL and EEF source enumeration criteria for the nonuniform noise case. The performances of the proposed DOA estimator and the proposed source enumeration criteria are studied through numerical simulations.

The remainder of this paper is organized as follows. The signal model is described in section II. The stochastic maximum likelihood algorithm for the nonuniform noise case and the approach that is proposed in this paper for calculating the SML estimate of model parameters is discussed in section III. In section IV, we discuss the SML source enumeration criteria AIC, MDL, and EEF for the nonuniform noise case. In section V, the approaches proposed for estimating number of sources with these information criteria are discussed and the performance of the proposed DOA estimation and source enumeration approaches are investigated using computer simulations. Finally, section VI concludes the paper.

## II. SIGNAL MODEL

Consider an array composed of $M$ sensors receiving $q$ ( $q < M$ ) narrowband signals from far-field sources. We assume that the array and the sources lie in the same plane. The array output $\mathbf{x}(t)$ at time instant $t$ can be modeled as

$$\mathbf{x}(t) = \mathbf{A}(\psi)\mathbf{s}(t) + \mathbf{n}(t), \quad (1)$$

where

$$\mathbf{A}(\psi) = [\mathbf{a}(\psi_1), \mathbf{a}(\psi_2), ..., \mathbf{a}(\psi_q)] \quad (2)$$

is the $M \times q$ matrix containing the array steering vectors $\mathbf{a}(\psi_i)$, $i = 1,2,...,q$, $\psi = [\psi_1,...,\psi_q]^T$ is the vector of source DOAs, $\mathbf{s}(t)$ is the $q \times 1$ vector of source signals, $\mathbf{n}(t)$ is the $M \times 1$ vector of temporally and spatially white noise, and $[.]^T$ stands for the transpose. It is assumed that the geometry of the array of sensors is such that $\mathbf{A}(\psi)$ is full rank for all source DOAs of interest. As a result, $\mathbf{A}^H(\psi)\mathbf{A}(\psi)$ will always be positive definite.

Using (1) and under the assumption that noise is uncorrelated with the source signals, the covariance matrix of the array output can be written as

$$\mathbf{R} = E\{\mathbf{x}(t)\mathbf{x}^H(t)\} = \mathbf{A}(\psi)\mathbf{P}\mathbf{A}^H(\psi) + \mathbf{Q}, \quad (3)$$

where $E\{.\}$ and $(.)^H$ stand for expectation and Hermitian transpose respectively, and $\mathbf{P}$ and $\mathbf{Q}$ are the $q \times q$ signal covariance matrix and the $M \times M$ noise covariance matrix defined by

$$\mathbf{P} = E\{\mathbf{s}(t)\mathbf{s}^H(t)\}, \quad (4)$$

$$\mathbf{Q} = E\{\mathbf{n}(t)\mathbf{n}^H(t)\}. \quad (5)$$

In the presence of nonuniform noise, the covariance matrix of noise is a diagonal matrix of the form

$$\mathbf{Q} = \text{diag}\{\sigma_1^2, \sigma_2^2, ..., \sigma_M^2\}, \quad (6)$$

where diag$\{.\}$ denotes a diagonal matrix and $\sigma_m^2$ is the noise power of the $m$th sensor.

### III. THE STOCHASTIC MAXIMUM LIKELIHOOD ALGORITHM

Under the assumption that the snapshots are independent and identically distributed zero-mean complex Gaussian random variables, the log-likelihood function for the array outputs $\{\mathbf{x}(t)\}_{t=1}^N$ is given by [13], [15], and [16],

$$L(\boldsymbol{\theta}) = L(\psi, \mathbf{P}, \mathbf{Q}) = -NM \ln(\pi) - N \ln[\det(\mathbf{R})] \\ - N \text{tr}(\mathbf{R}^{-1}\hat{\mathbf{R}}), \quad (7)$$

where ln(.), det(.) and tr(.) denote logarithm, determinant and trace respectively, $\boldsymbol{\theta}$ is the vector containing all parameters of $\psi$, $\mathbf{P}$ and $\mathbf{Q}$ (i.e., the vector containing all model parameters), $N$ is the number of snapshots, $\mathbf{R}$ is given by (3), and $\hat{\mathbf{R}}$ is the sample covariance matrix

$$\hat{\mathbf{R}} = \frac{1}{N}\sum_{t=1}^{N}\mathbf{x}(t)\mathbf{x}^H(t). \quad (8)$$

The maximum likelihood estimates for parameters of $\psi$, $\mathbf{P}$ and $\mathbf{Q}$ are obtained by maximizing (7) with respect to these parameters. Equivalently, these estimates can be obtained by minimizing the function

$$L'(\boldsymbol{\theta}) = L'(\psi, \mathbf{P}, \mathbf{Q}) = \ln[\det(\mathbf{R})] + \text{tr}(\mathbf{R}^{-1}\hat{\mathbf{R}}). \quad (9)$$

Following a similar derivation as [13], the optimum $\mathbf{P}$ that minimizes (9) in the presence of nonuniform noise is obtained as

$$\hat{\mathbf{P}} = (\tilde{\mathbf{A}}^H\tilde{\mathbf{A}})^{-1}\tilde{\mathbf{A}}^H\hat{\tilde{\mathbf{R}}}\tilde{\mathbf{A}}(\tilde{\mathbf{A}}^H\tilde{\mathbf{A}})^{-1} - (\tilde{\mathbf{A}}^H\tilde{\mathbf{A}})^{-1}, \quad (10)$$

where

$$\tilde{\mathbf{A}} = Q^{-\frac{1}{2}}\mathbf{A}, \quad (11)$$

$$\hat{\tilde{\mathbf{R}}} = Q^{-\frac{1}{2}}\hat{\mathbf{R}}Q^{-\frac{1}{2}}. \quad (12)$$

Note that we have dropped the argument $\psi$ in $\mathbf{A}$ and $\tilde{\mathbf{A}}$ for simplicity. The minimum value of $L'(\psi, \mathbf{P}, \mathbf{Q})$ which corresponds to (10) is

$$L'(\psi, \hat{\mathbf{P}}, \mathbf{Q}) = \ln[\det(\mathbf{Q})] + \ln[\det(\tilde{\mathbf{A}}^H\hat{\tilde{\mathbf{R}}}\tilde{\mathbf{A}}(\tilde{\mathbf{A}}^H\tilde{\mathbf{A}})^{-1})] \\ + \text{tr}[(\mathbf{I}_M - \mathbf{P}_{\tilde{\mathbf{A}}})\hat{\tilde{\mathbf{R}}}] + q, \quad (13)$$

where $\mathbf{I}_M$ is the $M \times M$ identity matrix, and

$$\mathbf{P}_{\tilde{\mathbf{A}}} = \tilde{\mathbf{A}}(\tilde{\mathbf{A}}^H\tilde{\mathbf{A}})^{-1}\tilde{\mathbf{A}}^H. \quad (14)$$

The SML estimates for $\mathbf{Q}$ and $\psi$ can be obtained by minimizing (13) with respect to the parameters of $\mathbf{Q}$ and $\psi$. To concentrate $L'(\psi, \hat{\mathbf{P}}, \mathbf{Q})$ further, one would fix $\psi$ and find an estimator of $\mathbf{Q}$ that minimizes $L'(\psi, \hat{\mathbf{P}}, \mathbf{Q})$ [11]. Unfortunately, a closed-form separable estimator of **Q** that minimizes (13) seems to be analytically unavailable and this makes further simplification impossible [11]. However, two estimators for **Q** in the presence of nonuniform noise have been recently proposed in [12] and [14]. By inserting each of these estimates of **Q** in (13) and then searching for $\psi$, one can obtain an estimate of $\psi$. In addition, by inserting these estimates of **Q** and $\psi$ in (10), one can obtain an estimate of **P**. This approach makes it possible for us to obtain SML estimates for source DOAs in the presence of nonuniform noise. Furthermore, this approach gives us an estimate of stochastic maximum-likelihood function for the nonuniform noise case and makes it possible for us to realize stochastic maximum likelihood AIC, MDL and EEF source enumeration criteria for the nonuniform noise case.

The iterative ML subspace estimation (IMLSE) algorithm proposed in [12] estimates the matrices **B** and **Q** by maximizing the stochastic maximum likelihood function in an iterative manner, where

$$\mathbf{P} = \mathbf{L}\mathbf{L}^H, \tag{15}$$
$$\mathbf{B} = \mathbf{A}\mathbf{L}, \tag{16}$$

where **L** and **B** are $q \times q$ and $M \times q$ matrices, respectively, and

$$\mathbf{R} = \mathbf{A}\mathbf{P}\mathbf{A}^H + \mathbf{Q} = \mathbf{B}\mathbf{B}^H + \mathbf{Q}. \tag{17}$$

The iteration continues until convergence. The algorithm proposed in [14] is a noniterative method that uses eigendecomposition to estimate **Q** and noise subspace. This algorithm has a computational complexity much less than that of IMLSE. As we discussed above, we will use the estimates of **Q** obtained by both of these algorithms to estimate source DOAs by the SML algorithm. Furthermore, we will use the estimates of **Q** obtained by both of these algorithms to determine the number of sources by source enumeration criteria.

## IV. SOURCE ENUMERATION METHODS

Assume that $\hat{\psi}_q$, $\hat{\mathbf{P}}_q$, and $\hat{\mathbf{Q}}_q$ are the stochastic ML estimates of $\psi$, **P,** and **Q** respectively, under the assumption that the number of signal sources is equal to $q$ and the noise is spatially white and nonuniform. In the previous section, we discussed how one can calculate these estimates with admissible computational complexity. Therefore, we can use these estimates to obtain and realize several source enumeration criteria with admissible computational complexity. The well-known model order selection criteria AIC and MDL are defined as

$$\text{AIC}(q) = NL'(\hat{\boldsymbol{\theta}}_q) + k_q = NL'(\hat{\psi}_q, \hat{\mathbf{P}}_q, \hat{\mathbf{Q}}_q) + k_q, \tag{18}$$

$$\hat{q}_{\text{AIC}} = \arg\min_q \{\text{AIC}(q)\}, \tag{19}$$

$$\text{MDL}(q) = NL'(\hat{\boldsymbol{\theta}}_q) + \frac{1}{2}k_q \ln(N)$$
$$= NL'(\hat{\psi}_q, \hat{\mathbf{P}}_q, \hat{\mathbf{Q}}_q) + \frac{1}{2}k_q \ln(N), \tag{20}$$

$$\hat{q}_{\text{MDL}} = \arg\min_q \{\text{MDL}(q)\}, \tag{21}$$

where $\hat{\boldsymbol{\theta}}_q$ is the vector containing all parameters of $\hat{\psi}_q$, $\hat{\mathbf{P}}_q$, and $\hat{\mathbf{Q}}_q$, $k_q$ is the number of free real parameters characterizing the stochastic signals model [4], $\hat{q}_{\text{AIC}}$ and $\hat{q}_{\text{MDL}}$ are the number of sources chosen by the AIC and MDL criteria respectively, and it can be seen from (9) and (3) that

$$L'(\hat{\boldsymbol{\theta}}_q) = L'(\hat{\psi}_q, \hat{\mathbf{P}}_q, \hat{\mathbf{Q}}_q) = \ln[\det(\hat{\mathbf{R}}_q)] + \text{tr}(\hat{\mathbf{R}}_q^{-1}\hat{\mathbf{R}}), \tag{22}$$

$$\hat{\mathbf{R}}_q = \mathbf{A}(\hat{\psi}_q)\hat{\mathbf{P}}_q\mathbf{A}^H(\hat{\psi}_q) + \hat{\mathbf{Q}}_q. \tag{23}$$

The parameter $k_q$ is equal to the total number of free real parameters in $\psi$, **P,** and **Q**. The number of the total free real parameters in **P** and $\psi$ is equal to $q^2 + q$ [4]. In fact, it can be seen that the number of free real parameters in **P** is equal to $q^2$, and the number of free real parameters in $\psi$ is equal to $q$. In addition, in presence of spatially white nonuniform noise, the number of free real parameters in **Q** is equal to $M$. Consequently, we have

$$k_q = q^2 + q + M. \tag{24}$$

The exponentially embedded family (EEF) criterion for model order selection was first proposed in [17]. In [3] a version of the EEF criterion for source enumeration is proposed but cannot be extended to the nonuniform noise problem here because the SML model is not considered in that version. Therefore, we derive the SML-based version of the EEF criterion for source enumeration in the presence of nonuniform noise. The EEF criterion selects the number of sources that maximizes

$$\text{EEF}(q) = \left\{ L_{G_q}(\mathbf{X}) - k_q \left[ \ln\left(\frac{L_{G_q}(\mathbf{X})}{k_q}\right) + 1 \right] \right\} \times u\left(\frac{L_{G_q}(\mathbf{X})}{k_q} - 1\right), \quad (25)$$

that is,

$$\hat{q}_{\text{EEF}} = \arg\max_q \{\text{EEF}(q)\}, \quad (26)$$

where $\hat{q}_{\text{EEF}}$ is the number of sources chosen by the EEF criterion, $u(.)$ is the unit step function defined as

$$u(t) = \begin{cases} 1 & ; t \geq 0 \\ 0 & ; t < 0 \end{cases}, \quad (27)$$

and $L_{G_q}(\mathbf{X})$ is defined as

$$L_{G_q}(\mathbf{X}) = 2[L(\hat{\boldsymbol{\theta}}_q) - L(\hat{\boldsymbol{\theta}}_0)] = -2N[L'(\hat{\boldsymbol{\theta}}_q) - L'(\hat{\boldsymbol{\theta}}_0)], \quad (28)$$

where

$$\mathbf{X} = [\mathbf{x}(1), \mathbf{x}(2), ..., \mathbf{x}(N)], \quad (29)$$

and $\hat{\boldsymbol{\theta}}_0$ is the vector of estimated model parameters under the assumption that there are no signal sources impinging on the array of sensors ($q=0$).

It can be seen from (25)-(28) that in order to obtain $\hat{q}_{\text{EEF}}$, in addition to calculating $L'(\hat{\boldsymbol{\theta}}_q)$, we need to calculate $L'(\hat{\boldsymbol{\theta}}_0)$. In the case that $q=0$, we have

$$\mathbf{R} = \mathbf{Q} = \text{diag}\{\sigma_1^2, \sigma_2^2, ..., \sigma_M^2\}. \quad (30)$$

Substituting (30) into (9) we get

$$L'(\boldsymbol{\theta}) = \ln(\sigma_1^2 \sigma_2^2 ... \sigma_M^2) + \text{tr}\left(\text{diag}\left\{\frac{1}{\sigma_1^2}, \frac{1}{\sigma_2^2}, ..., \frac{1}{\sigma_M^2}\right\} \hat{\mathbf{R}}\right)$$

$$= \sum_{i=1}^{M} \ln(\sigma_i^2) + \sum_{i=1}^{M} \frac{1}{\sigma_i^2} \hat{\mathbf{R}}(i,i), \quad (31)$$

where $\hat{\mathbf{R}}(i,i)$ is the $i$th element on the main diagonal of $\hat{\mathbf{R}}$. Now, we should maximize (31) with respect to $\{\sigma_i^2\}_{i=1}^{M}$. Taking the partial derivative of $L'(\boldsymbol{\theta})$ with respect to each $\sigma_i^2$, we get

$$\frac{\partial L'(\boldsymbol{\theta})}{\partial \sigma_i^2} = \frac{1}{\sigma_i^2} - \frac{1}{\sigma_i^2} \hat{\mathbf{R}}(i,i) \quad ; i = 1, 2, ..., M. \quad (32)$$

Setting (32) equal to zero, we obtain

$$\hat{\sigma}_i^2 = \hat{\mathbf{R}}(i,i) \quad ; i = 1, 2, ..., M. \quad (33)$$

Substituting (33) into (31), we obtain

$$L'(\hat{\boldsymbol{\theta}}_0) = \sum_{i=1}^{M} \hat{\mathbf{R}}(i,i) + M. \quad (34)$$

Thus, we use (22) and (34) to obtain the value of $L_{G_q}(\mathbf{X})$ from (28) for each $q$. Finally, substituting these values of $L_{G_q}(\mathbf{X})$ into (25) and (26), we obtain $\hat{q}_{\text{EEF}}$. Note that the value of $L'(\hat{\boldsymbol{\theta}}_0)$ given by (34) is also used in (18) and (20) for obtaining AIC(0) and MDL(0), respectively.

## V. SIMULATION RESULTS

In this section, the performances of the nonuniform SML DOA estimation method and the SML-based source enumeration methods proposed in previous sections are demonstrated by using simulations.

### A. DOA Estimation

In this subsection, we will test and compare the performances of four nonuniform ML DOA estimation algorithms. Two of these algorithms are two versions of the nonuniform SML algorithm discussed in section III with noise covariance matrix estimate $\hat{\mathbf{Q}}$ calculated by either the IMLSE algorithm of [12] or the noniterative algorithm of [14]. The other two algorithms are the two versions of the nonuniform deterministic maximum-likelihood (DML) algorithm with noise covariance matrix estimate $\hat{\mathbf{Q}}$ calculated by either the IMLSE algorithm of [12] or the noniterative algorithm of [14]. Note that in order to reduce the computational complexity, the SML and DML algorithms are approximated in all of the simulations of this paper by the alternating maximization (AM) and the alternating projection (AP) algorithms respectively.

A uniform linear array (ULA) composed of $M$=6 sensors, which are separated by half wavelength, is considered. Two equal-power narrowband sources impinge on the array. The signal to noise ratio (SNR) for each source is defined as

$$\text{SNR} = \frac{\sigma_s^2}{M} \sum_{i=1}^{M} \frac{1}{\sigma_i^2}, \qquad (35)$$

where $\sigma_s^2$ denotes the signal power for each source. The worst noise power ratio (WNPR) is defined as

$$\text{WNPR} = \frac{\sigma_{\max}^2}{\sigma_{\min}^2}, \qquad (36)$$

where $\sigma_{\min}^2$ and $\sigma_{\max}^2$ are the lowest and the highest sensor noise power in the array, respectively. The noise covariance matrix in all simulations is as
$$\mathbf{Q} = \text{diag}\{9, 1, 25, 0.25, 6.25, 25\}. \qquad (37)$$

As a result, WNPR is equal to 100. In each figure of this subsection the root mean squared errors (RMSEs) of DOA estimates are plotted. The RMSE of DOA estimates is defined as

$$\text{RMSE} = \sqrt{\frac{1}{Kq} \sum_{k=1}^{K} \sum_{l=1}^{q} (\hat{\psi}_{k,l} - \psi_l)^2}, \qquad (38)$$

where $\psi_l$ is the DOA of the $l$th source, $\hat{\psi}_{k,l}$ is the estimate of $\psi_l$ in the $k$th simulation run, $q$ is the number of sources, and $K$ is the number of independent simulation runs. In addition, the stochastic and deterministic Cramer-Rao bounds (CRBs) [10] are plotted in each figure. The number of independent simulation runs is set to $K$=400 in all figures.

In the first scenario, we consider two uncorrelated signal sources with DOAs set to $\psi_1 = -3°$ and $\psi_2 = 4°$. The number of snapshots is set to $N$=300. Fig. 1 shows the RMSEs of DOA estimates for the tested methods versus SNR. It can be seen from this figure that the methods which use the noniterative algorithm of [14] for estimating $\mathbf{Q}$ have better performance in high SNRs. In low and high SNR values, the performance of SML-based methods is slightly better than that of the corresponding DML-based methods, but in middle SNR values, the situation is the reverse.

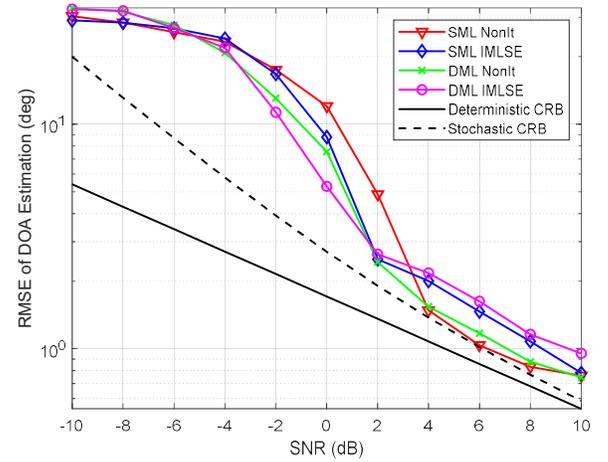

Fig. 1. RMSEs of DOA estimates versus SNR for two uncorrelated sources.

In the second scenario, we consider two correlated signal sources with correlation coefficient set to 0.95. Other settings are the same as the first scenario. Fig. 2 shows the RMSEs of DOA estimates for the tested methods versus SNR. It can be seen from this figure that the methods which use the noniterative algorithm of [14] for estimating $\mathbf{Q}$ give very poor DOA estimates. This is not surprising because the algorithm proposed in [14] uses the assumption that the sources are uncorrelated. It can also be seen from this figure that the SML-based method that uses the IMLSE algorithm of [12] for estimating $\mathbf{Q}$ has the best performance in all SNRs. Specifically, note that the performance of this method is much better than that of the DML-based approach recently proposed in [12].

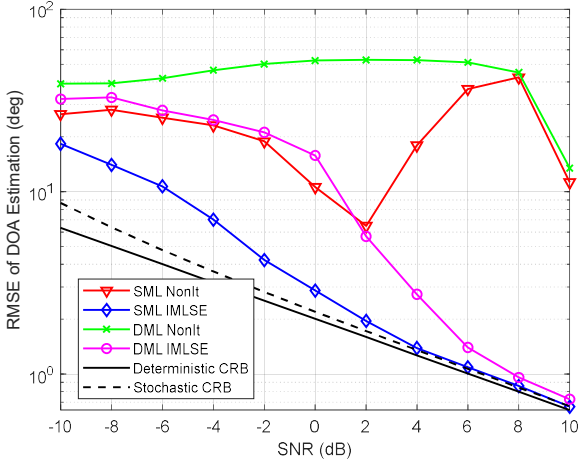

Fig. 2. RMSEs of DOA estimates versus SNR for two correlated sources.

In the third scenario, we consider two uncorrelated signal sources and compare the behaviors of DOA estimation methods in different source angle separations. The SNR is set to 0 dB, the number of snapshots is set to $N=300$, the DOA of the first source is set to $\psi_1 = -10°$, and $\psi_2$ varies from $-8°$ to $10°$. Fig. 3 shows the RMSEs of DOA estimates for the tested methods versus angle separation. It can be seen from this figure that the methods which use the noniterative method of [14] for estimating $\mathbf{Q}$ have better performance in high angle separations.

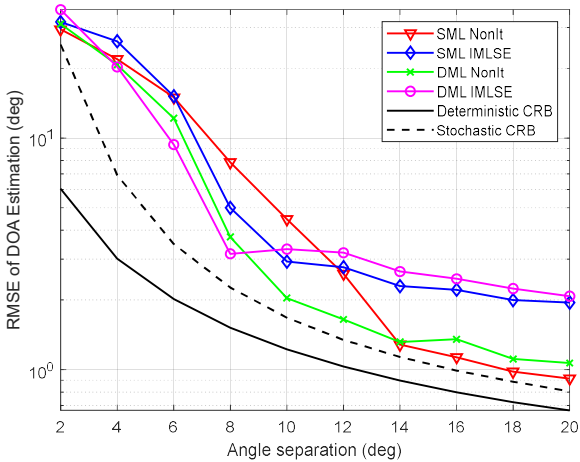

Fig. 3. RMSEs of DOA estimates versus angle separation for two uncorrelated sources.

On the whole, we can say that the ML methods which use the noniterative algorithm of [14] for estimating $\mathbf{Q}$ have better performance in high SNRs and/or high angle separations. However, these methods have very poor performance in the presence of correlated sources. In addition, the important advantage of the SML-based methods over the DML-based methods is that they are more robust against the correlation of sources.

*B. Source Enumeration*

In this subsection, we will test and compare the performances of the nonuniform SML-based AIC, MDL, and EEF source enumeration criteria discussed in section IV. In all of these three information criteria, we need to calculate $L'(\hat{\boldsymbol{\theta}}_q)$ for each source number $q$. We will propose three approaches for calculating $L'(\hat{\boldsymbol{\theta}}_q)$ and the performances of these three approaches are compared in this subsection. The first approach uses the IMLSE algorithm of [12] to calculate the estimates $\hat{\mathbf{Q}}_q$ and $\hat{\mathbf{B}}_q$ for each source number $q$ and then inserts these estimates in (17) to obtain the estimate $\hat{\mathbf{R}}_q$. At last, (22) is used for calculating $L'(\hat{\boldsymbol{\theta}}_q)$. The second approach uses the IMLSE algorithm of [12] to calculate the estimate $\hat{\mathbf{Q}}_q$ for each source number $q$ and then uses the nonuniform SML algorithm discussed in section III to obtain the estimates $\hat{\boldsymbol{\psi}}_q$ and $\hat{\mathbf{P}}_q$. Then, by inserting $\hat{\mathbf{Q}}_q$, $\hat{\boldsymbol{\psi}}_q$, and $\hat{\mathbf{P}}_q$ in (23) and (22), $L'(\hat{\boldsymbol{\theta}}_q)$ is calculated. The third approach is the same as the second approach except that the estimate $\hat{\mathbf{Q}}_q$ is calculated using the noniterative algorithm of [14].

In all simulation scenarios of this subsection, the array of sensors and the noise covariance matrix $\mathbf{Q}$ are the same as the previous subsection and two equal-power narrowband sources are impinging on the array. In addition, the number of independent simulation runs and the number of snapshots are set to $K=100$ and $N=100$, respectively. In all simulation scenarios the minimum and maximum source numbers that can be chosen by the source enumeration criteria are set to $q=0$ and $q=M-1$, respectively.

In the first scenario of this subsection, we consider two uncorrelated narrowband signal sources with DOAs set to $\psi_1 = -5°$ and $\psi_2 = 6°$. Fig. 4 shows the number of successful source enumerations for the tested criteria with either of the three approaches of estimating $L'(\hat{\boldsymbol{\theta}}_q)$, versus SNR. It can be seen from this figure that when the first or the third approach of estimating $L'(\hat{\boldsymbol{\theta}}_q)$ is used, the performance of MDL and EEF criteria is better than that of AIC in high SNRs, and the performance of AIC and EEF criteria is better than that of MDL in low SNRs. Furthermore, when the second approach of estimating $L'(\hat{\boldsymbol{\theta}}_q)$ is exploited, the performance of all three information criteria is poor in low SNRs. It is worth mentioning that for all three approaches, the performance of EEF is similar to that of MDL in high SNRs and similar to that of AIC in low SNRs.

In the second scenario, we consider two correlated narrowband signal sources with correlation coefficient set to 0.95. Other settings are the same as the first scenario. Fig. 5 shows the number of successful source enumerations for the tested criteria with either of the three approaches of estimating $L'(\hat{\boldsymbol{\theta}}_q)$, versus SNR. It can be seen from this figure that, in spite of the correlation between source signals, the three criteria still have good performance in high SNRs if the second approach is used for estimating $L'(\hat{\boldsymbol{\theta}}_q)$. In other words, the three order selection criteria are robust against correlation of source signals if the second approach is used for estimating $L'(\hat{\boldsymbol{\theta}}_q)$. In fact, the performance of the three criteria with the second approach is even better than that of Fig. 4. It can also be seen from this figure that the performance of the three criteria with the third approach is very poor in high SNRs in this case. Furthermore, the performance of the MDL and EEF criteria with the first approach is poor in many SNR values.

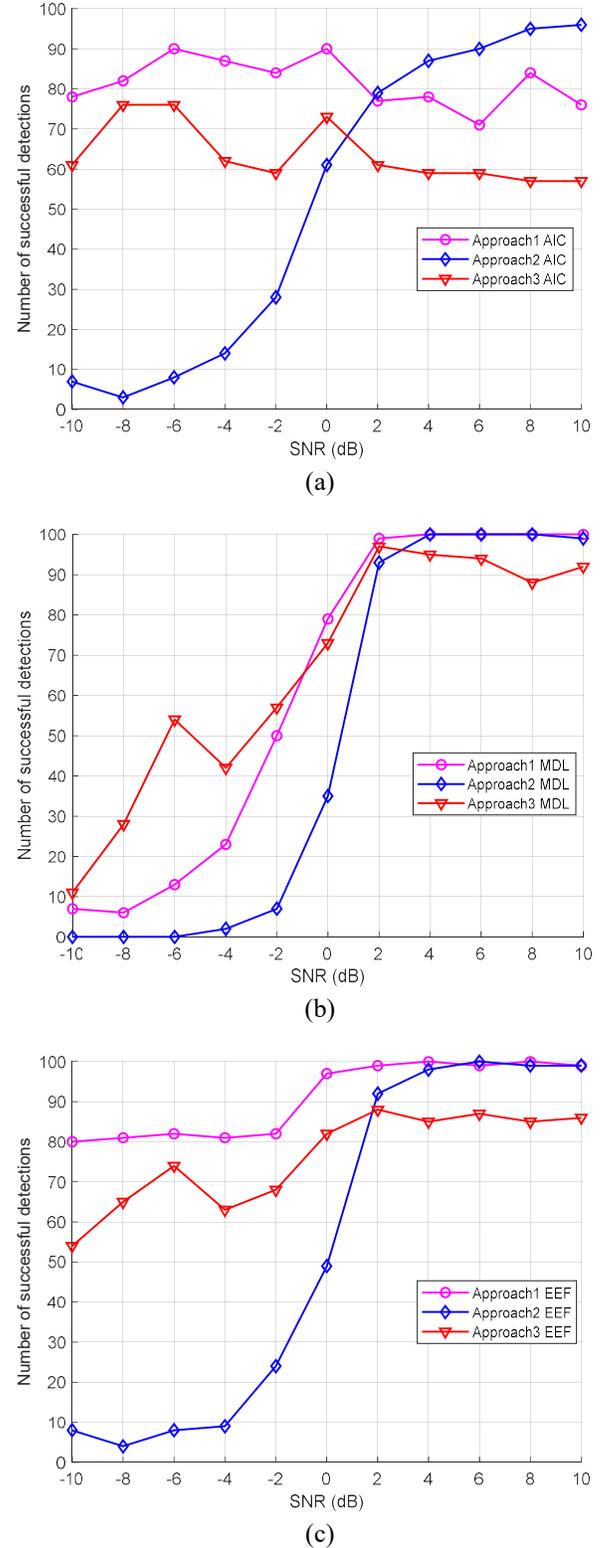

Fig. 4. Number of successful source enumerations versus SNR with either of the three approaches for the criteria (a) AIC (b) MDL and (c) EEF, when two uncorrelated sources impinge on the array.

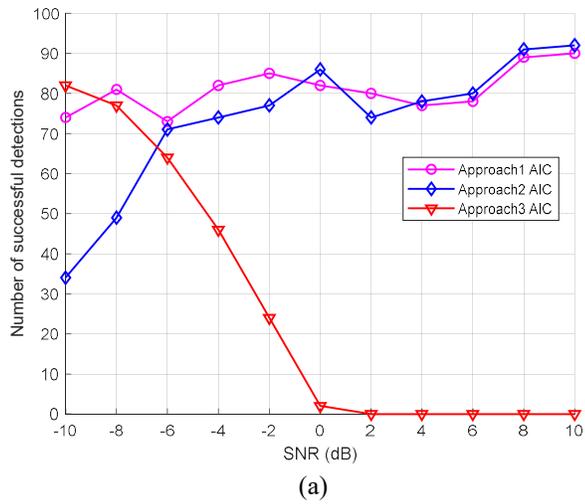

(a)

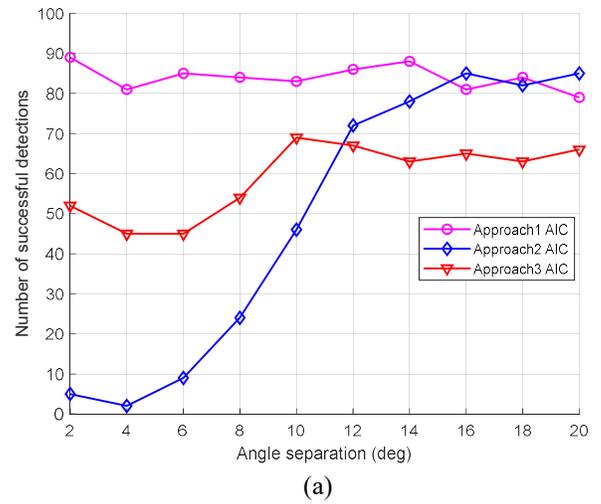

(a)

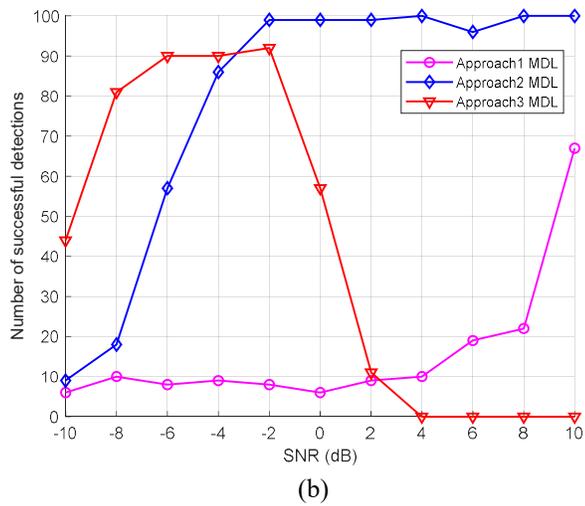

(b)

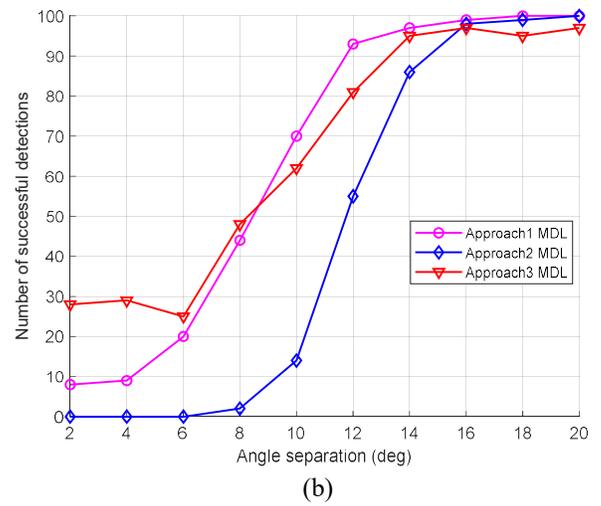

(b)

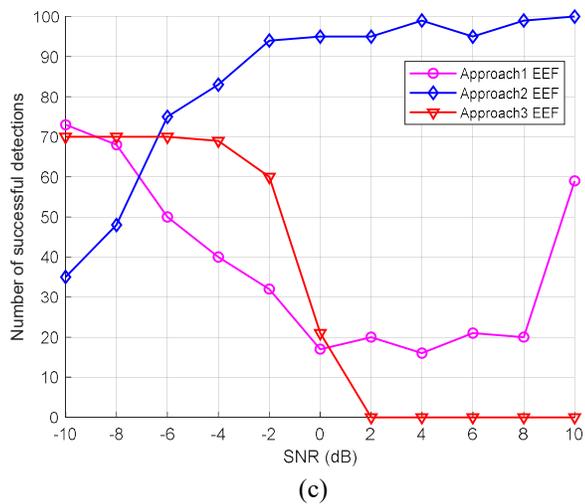

(c)

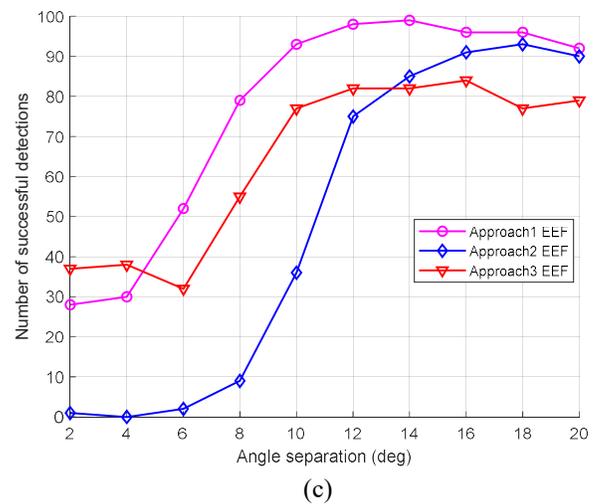

(c)

Fig. 5. Number of successful source enumerations versus SNR with either of the three approaches for the criteria (a) AIC (b) MDL and (c) EEF, when two correlated sources impinge on the array.

Fig. 6. Number of successful source enumerations versus angle separation with either of the three approaches for the criteria (a) AIC (b) MDL and (c) EEF, when two uncorrelated sources impinge on the array.

In the third scenario, we consider two uncorrelated signal sources and compare the behaviors of source enumeration methods in different source angle separations. The SNR is set to 0 dB, the DOA of the first source is set to $\psi_1 = -10°$, and $\psi_2$ varies from $-8°$ to $10°$. Fig. 6 shows the number of successful source enumerations for the tested criteria with either of the three approaches of estimating $L'(\hat{\boldsymbol{\theta}}_q)$, versus angle separation. It can be seen from this figure that, in all three approaches, the criteria MDL and EEF have poor performance in small angle separations and good performance in large angle separations. However, if the first or the third approach is used, all the three criteria show better performance in small angle separations.

## VI. CONCLUSION

In spite of the uniform noise case, concentration of the stochastic likelihood function with respect to the diagonal elements of noise covariance matrix is not possible in the presence of nonuniform noise. Therefore, using this function for calculating the SML estimates of the source DOAs, the noise covariance matrix, and the signal covariance matrix, needs a multidimensional search with very high computational complexity. The information criteria AIC, MDL, and EEF need these estimates of signal model parameters to estimate the number of source signals. Therefore, source enumeration with these criteria also needs very high computational complexity in the nonuniform noise case. In this paper, we proposed new approaches for solving this problem and estimating source DOAs and source number in nonuniform noise case with admissible complexity.

Computer simulations were used to demonstrate the performance of the proposed approaches in various scenarios. Simulation results show that some of the proposed SML-based DOA estimation and source enumeration approaches are robust against high correlation between source signals. In the case of high correlation between source signals, the proposed SML-based DOA estimation method that uses the IMLSE algorithm of [12] for estimating **Q** has the best performance in all SNRs among all methods including the DML-based approach recently proposed in [12].

In the case that the signal sources are uncorrelated, the source enumeration information criteria usually have the best performance in low SNRs and/or low angle separations with the first and the third approaches of estimating $L'(\hat{\boldsymbol{\theta}}_q)$. On the other hand, the second approach for source enumeration is robust against high correlation between source signals. In the case that the signal sources are highly correlated, the source enumeration information criteria usually have the best performance in middle and high SNRs with the second approach.